\documentclass[12pt]{article}

\usepackage{lscape}
\usepackage{multirow}
\usepackage{amssymb,amsmath}
\usepackage{amstext}
\usepackage{amscd}
\usepackage{slashed}
\usepackage{graphics}
\usepackage{epsfig}

\def\wt{\widetilde}
\newcommand{\Rho}{{\mbox{\sf P}}}

\def\m{\mu}

\def\be{\begin{equation}}
\def\ee{\end{equation}}
\def\beq{\begin{equation}}
\def\eeq{\end{equation}}
\def\bea{\begin{eqnarray}}
\def\eea{\end{eqnarray}} 

\def\eqn#1{(\ref{#1})}

\def\nn{\nonumber}

\def\sideremark#1{\ifvmode\leavevmode\fi\vadjust{\vbox to0pt{\vss
 \hbox to 0pt{\hskip\hsize\hskip1em
 \vbox{\hsize3cm\tiny\raggedright\pretolerance10000
  \noindent #1\hfill}\hss}\vbox to8pt{\vfil}\vss}}}

\begin{document}
\thispagestyle{empty}

%\vspace{.6cm}
\setcounter{footnote}{0}
\begin{center}
\vspace{-40mm}
{\Large
 {\bf Weyl's  Gauge Invariance:}\\[2mm] {\bf Conformal~Geometry,  Spinors, Supersymmetry, and Interactions }\\[4mm]

 {\sc \small
     A.~Shaukat$^{\mathfrak S}$,  and A.~Waldron$^{\mathfrak W}$\\[2mm]
{
    %        {\em\small ${}^{\mathfrak G}\!$
      %     Department of Mathematics\\ 
       %    The University of Auckland,
       %     Auckland, New Zealand\\
        %    {\tt gover@math.auckland.ac.nz}\\[1mm]  
            ${}^{\mathfrak S}\!$
            Physics Department\\  
            University of California,
            Davis CA 95616, USA\\
            {\tt ashaukat@ucdavis.edu}
            \\[1mm] 
            ${}^{\mathfrak W}\!$
            Department of Mathematics\\ 
            University of California,
            Davis CA 95616, USA\\[-2mm]
            {\tt wally@math.ucdavis.edu}
            }}\\[-2mm]

 }

\vspace{6mm}

{\sc Abstract}\\[-20mm]
\end{center}

\vspace{-5mm}

{\small
\begin{quote}
We extend our program, of coupling theories to scale in order to make their Weyl invariance manifest, to include interacting theories, fermions and supersymmetric theories. The results produce mass terms coinciding with the standard ones
for universes that are Einstein, but are novel in general backgrounds. They are generalizations of the gravitational couplings of a conformally improved scalar  to fields with general scaling and tensor properties. The couplings we find are more general than just trivial ones
following from the conformal compensating mechanisms. In particular, in the setting where a scale gauge field (or dilaton) is
included, masses correspond to Weyl weights of fields organized in ``tractor''  multiplets. Breitenlohner--Freedman bounds follow directly from 
reality of these weights. Moreover, massive, massless and partially massless theories are handled in a uniform framework. Also, {\it bona fide} Weyl invariant 
theories (invariant without coupling to scale) can be directly derived in this approach. The results are based on the tractor calculus
approach to conformal geometry, in particular we show how to handle fermi fields, supersymmetry and Killing spinors using tractor techniques.
Another useful consequence of the construction is that it automatically produces the (anti) de Sitter theories obtained by log-radial reduction of 
Minkowski theories in one higher dimension. Theories  presented in detail include interacting scalars, spinors, Rarita--Schwinger fields, and
the interacting Wess--Zumino model. 
\end{quote}
}

\tableofcontents

\section{Introduction}

Weyl invariance is often regarded as a symmetry principle obeyed by particular classes of theories. In a recent series of
papers~\cite{ Gover:2008sw,Gover:2008pt}, we argued that, in fact, it should be viewed in the same manner as general coordinate invariance:
{\it all} \   theories should respect Weyl invariance.   As a consequence, physical quantities are then independent of any local choice of unit system.  
Just as one couples to a metric to ensure diffeomorphism invariance, when necessary, theories should be coupled to scale (a --not necessarily dynamical-- dilaton
field) to guarantee Weyl invariance. Moreover, analogous to the importance of finding all possible diffeomorphism invariant theories, a crucial
problem is to find all possible Weyl and diffeomorphism invariant theories. The solution to this problem not only gives a simple way to find those 
peculiar theories that are invariant without coupling to scale, but also leads to deep insights into many aspects of those that are not.

Ultimately, we believe that this program will lead to a deeper understanding of quantum effects, in particular the 
renormalization group and AdS/CFT correspondence, especially from a holographic  renormalization group viewpoint~\cite{de Boer:1999xf}.
In the current paper, however, we restrict ourselves to a classical analysis and extend our previous results to (i) interactions, (ii) 
fermions and (iii) supersymmetry. The Paper is structured as follows: in the next Section, we discuss how to couple scalar fields to scale to as well as gravity.
Following that,  we review
the tractor technology required to handle spinors and, in Section~\ref{Tractor Dirac Equation}, 
use it to  formulate massive and massless spin 1/2 particles in a single, Weyl invariant framework.  Extending these ideas,  in Section~\ref{Tractor Rarita--Schwinger Equation} we couple the spin 3/2  Rarita--Schwinger equation to scale. We then combine bose and fermi
models in Section~\ref{Supersymmetry} and describe supersymmetric theories, concentrating on the Wess--Zumino model. Section~\ref{Interactions}
describes an interacting Wess--Zumino model coupled to scale.  Since tractors are intimately related to the projective approach to conformal geometry,  they naturally produce the log-radial reduction procedure~\cite{Biswas:2002nk,Hallowell:2005np} used to obtain constant curvature theories from flat space ones in a dimension higher. The log-radial reduction for spinors is described in Appendix~\ref{Doubled Reduction}.

\section{Coupling to Scale}

\begin{quote}
{\it My work always tried to unite the
Truth with the Beautiful, but when I
had to choose one or the other, I usually chose the Beautiful.}
\vspace{-.3cm}

\flushright{Hermann Weyl}
\end{quote}

\label{Coupling to Scale}
Classically, all physical theories are required to have a rigid scaling symmetry reflecting the freedom to globally choose any unit system  for a  given physical quantity. For example, if $x$ is a length, then $x + x^2$ cannot be a sensible answer to a physical question. The scaling properties of physical quantities are encoded by assigning them weights.
 Then, in an action principle, scale invariance  requires
\be
S[\Phi_i;\lambda_\alpha]=S[\Omega^{w_i}\Phi_i;\Omega^{w_\alpha}\lambda_\alpha]\, ,\label{units}
\ee
where $\Omega$ is a rigid parameter, $\{\Phi_i\}$ are the fields of the theory and $\{\lambda_\alpha\}$ are any dimensionful couplings.
In fact, only a single dimensionful coupling $\kappa$ (Newton's constant) is really needed, since all the others become dimensionless
upon multiplication by an appropriate power of~$\kappa$.

The equation~\eqn{units} simply says that physics is independent of the choice of unit system. Just as physical 
systems are required to be independent of {\it local} changes of coordinates (``diffeomorphisms''), a fundamental physical principle,
dating back to Weyl~\cite{Weyl:1918}, is that physics be independent of {\it local} choices of unit systems (``Weyl transformations'') as well. Generically,
diffeomorphism invariance is only achieved by introducing a gauge field -- the metric tensor~$g_{\mu\nu}$. (An exception to the rule, is
Chern--Simons theory, for example.) The same is true for Weyl invariance: generically a gauge field $\sigma$ is necessary although again
there are exceptions, notably the conformally improved wave equation, the massless Dirac equation and
Maxwell's equations in four dimensions. The gauge field $\sigma$ is often called a Weyl compensator or dilaton and was employed in this way
by Deser and Zumino~\cite{Zumino, Deser0}. We prefer to call $\sigma$ the scale, in concordance with the mathematics literature and also 
because of its geometric interpretation, which is simply as a local Newton's ``constant'' encoding how the choice of unit system varies over space and time. 

To efficiently study Weyl transformations, a conformal calculus is needed along the lines of tensor calculus for coordinate 
transformations. In~\cite{Gover:2008sw} we explained how the tractor calculus developed by conformal geometers~\cite{Thomas,BEG} provides exactly
such a calculus. We limit ourselves to a few key ideas here  and refer the reader to our papers~\cite{Gover:2008sw,Gover:2008pt} as well as the mathematical
literature~\cite{BEG,G, GP, CG} for further details. Under Weyl transformations, the metric and scale transform according to
\be\label{WT}
g_{\mu\nu}\mapsto \Omega^2 g_{\mu\nu}\, ,\qquad \sigma \mapsto \Omega \sigma\, , \qquad\Omega=\Omega(x)\, .
\ee
In $d$-dimensional tractor theory, fields are arranged in $\frak{so}(d,2)$ multiplets and, under Weyl transformations, transform
under particular $\frak{so}(d,2)$-valued ``tractor'' gauge transformations. A fundamental example of a weight zero tractor vector,
built only from the scale and the metric
is the {\it scale tractor}
\be\nn
I^M=\begin{pmatrix}\sigma \\[1mm] \partial^m \sigma \\[1mm] -\frac1d\, (\Delta + \Rho)\sigma\end{pmatrix}\, .
\ee
Here $M$ is an $\frak{so}(d,2)$ vector index, the index $\mu$ on $\partial_\mu$ has been flattened with an inverse vielbein
and the trace of the Schouten tensor $\Rho$ is proportional to the scalar curvature $R=2(d-1)\Rho$. 
Under Weyl transformations~\eqn{WT}, the scale tractor $I^M$ transforms as 
\be\nn
I^M \mapsto   U^M{}_N I^N \, , 
\ee 
where the $SO(d,2)$-valued matrix $U$ is given by 
\be\nn 
U=
\begin{pmatrix}
\Omega&0&\;0\;\\[2mm]
\Upsilon^m&\delta^m_n&\;0\;\\[2mm]
-\frac12\Omega^{-1}\,\Upsilon_r\Upsilon^r&-\Omega^{-1}\Upsilon_n&\Omega^{-1}
\end{pmatrix}\, , \qquad \Upsilon_\mu \equiv \Omega^{-1}\, \partial_\mu \Omega\, .
\ee 
Parabolic $SO(d,2)$ transformations of this special form are
 tractor gauge transformations. 

By virtue of the equivalence~\eqn{WT}, solving Einstein's equations means that we need only find a conformally Einstein metric.
Then we can arrange for the scale $\sigma$ to be constant and equal $\kappa^{\frac2{d-2}}$, which produces 
the actual Einstein metric from the conformal class of metrics. 
This amounts identically to requiring that the scale tractor be parallel with respect to the tractor connection defined on a weight
zero tractor $T^M$  by
\be\label{Dm}
{\cal D}_{\mu}\begin{pmatrix}T^+\\[1mm]T^m\\[1mm] T^{-}\end{pmatrix}\equiv
\begin{pmatrix}
\partial_\mu T^{+}-T_\mu\\[1mm]\nabla_\mu T^m +\Rho_\mu^m T^+ + e_\mu{}^m T^-\\[1mm]\partial_\mu T^- -\Rho_\mu^m T_m \end{pmatrix}\, ,
\ee
where the Schouten tensor is  the pure trace part of the Riemann tensor,
$R_{\mu\nu\rho\sigma}-W_{\mu\nu\rho\sigma}=4g_{[\mu[\rho}\Rho_{\nu]\sigma]}$. {\it I.e.}, $g_{\mu\nu}$ is conformally Einstein 
exactly when
\be\label{parallel}
{\cal D}_\mu I^M = 0\, .
\ee
Indeed, at the distinguished choice of scale, the left hand side gives
\be\nn
{\cal D}_{\mu} I^M\left|_{\sigma=\kappa^{\frac2{d-2}}}\right. = 
\kappa^{\frac {2} {d-2}} \begin{pmatrix}0\\[2mm]\Rho_\mu{}^m-\frac 1 d\, e_\mu{}^m\Rho \\[2mm] -\frac 1 d \,  \partial_\mu \Rho \end{pmatrix}\, ,
\ee
which verifies our claim.

The Einstein--Hilbert action also follows simply from the scale tractor; consider the manifestly Weyl invariant action
\be\label{TEH}
S(g_{\mu\nu},\sigma)=\frac{d(d-1)}2\int \frac{\sqrt{-g}}{\sigma^d}\,  I^M \eta_{MN} I^N=S(\Omega^2 g_{\mu\nu},\Omega\sigma)\, .
\ee
Here $\eta_{MN}={\tiny \begin{pmatrix}0&0&1\\0&\eta_{mn}&0\\1&0&0\end{pmatrix}}$ is the $\frak{so}(d,2)$ invariant metric.
At the canonical choice of scale we recover the Einstein--Hilbert action $S(g_{\mu\nu},\kappa^{\frac{2}{d-2}})=-\frac1{2\kappa^2}\int \!\sqrt{-g} \, R$.

To write physical theories in a Weyl invariant way, an operator taking the place of the Riemannian covariant derivative  in formulae is needed. It is provided by the
 Thomas $D$-operator which covariantly maps
weight $w$ tractors  to weight $w-1$ ones and is given by
\be\label{Thomas}
D^M\equiv\begin{pmatrix}(d+2w-2)w\\[2mm](d+2w-2){\cal
  D}^m\\[2mm]-({\cal D}_\nu{\cal D}^\nu + w \Rho)\end{pmatrix}. \ee 
Acting on the  weight one scale field $\sigma$, the Thomas $D$-operator  produces  the scale tractor 
$I^M = \frac{1}{d}D^M\sigma$. Although this operator is not Leibnitzian, its covariance under $\frak{so}(d,2)$ tractor gauge transformations
often makes it a more useful building block than the tractor covariant derivative~\eqn{Dm}  itself. 

Having presented the manifestly locally scale invariant, tractor, formulation of gravity, we now add matter fields and first focus on a
single scalar field~$\varphi$. The standard ``massless'' scalar field action $S=$\-$-\frac12 \int \sqrt{-g}$   $\nabla_\mu \varphi$   $g^{\mu\nu}$  $\nabla_\nu \varphi$ can easily be reformulated Weyl invariantly using the scale $\sigma$: From $\sigma$ we build the one-form
\be\nn
b=\sigma^{-1}d\sigma\, ,
\ee
with the simple transformation rule
\be\nn
b\mapsto b+\Upsilon\, .
\ee
Assigning the weight $w$ to the scalar~$\varphi$ 
\be\nn
\varphi\mapsto \Omega^w \varphi\, ,
\ee
then the  combination $\wt \nabla_\mu = \nabla_\mu - w b_\mu$ acting on $\varphi$ transforms covariantly
\be\nn
\wt \nabla_\mu \varphi\mapsto \Omega^w \wt \nabla_\mu \varphi\, .
\ee 
Hence we find an equivalent, but manifestly Weyl invariant, action principle
\be\label{trivial}
S=-\frac12\int  \frac{\sqrt{-g}}{\sigma^{d+2w-2}} \ \wt \nabla_\mu\varphi \, g^{\mu\nu}\,  \wt\nabla_\nu \varphi\, .
\ee
Of course, this is just the result of the ``compensating mechanism'', whereby, for any action involving a set of  
fields $\{\Phi_\alpha\}$ and their derivatives, replacing $\Phi_\alpha\mapsto \Phi_\alpha/\sigma^{w_\alpha}$
and $g_{\mu\nu}\mapsto g_{\mu\nu}/\sigma^2$ yields an equivalent Weyl invariant action. However, just as one 
searches for all possible coordinate invariant theories, we should also consider all possible locally scale invariant theories--theories independent
of local choices of unit system-- and then examine their physical consequences. In particular, the set of theories
invariant by virtue of the above Weyl compensator trick, does {\it not} map out the entire space of possible scalar
theories, even at the level of those quadratic in derivatives and fields. Tractor calculus is a very useful tool for such a study.

Motivated by ideas in conformal scattering theory~\cite{Scattering,Gscatter,GP}, in our recent work~\cite{Gover:2008pt} we proposed the Weyl invariant
theory with action principle
\be\nn
S=-\frac12\int  \frac{\sqrt{-g}}{\sigma^{d+2w-1}} \ \varphi\, I^M\! D_M \varphi \, .
\ee
Its difference from the action~\eqn{trivial} can also be expressed tractorially as 

\be \label{grav_mass_term}
\frac12\  w(d+w-1)\int   \frac{\sqrt{-g}}{\sigma^{d+2w}} \ \varphi\, I^M\! I_M \varphi\, ,
\ee 
which is reminiscent of the tractor Einstein--Hilbert action~\eqn{TEH}. 
Our proposal therefore amounted to coupling to the background geometry via the Weyl invariant\footnote{Note that since the Thomas $D$-operator is null ($D_MD^M=0$), $D_MI^M=0$. The weight one canonical tractor $X^M={\tiny\begin{pmatrix}0\\0\\1\end{pmatrix}}$ does not produce a new invariant either since
$X_M I^M=\sigma$. We also record the component form of the scale tractor $$I^M=\sigma\begin{pmatrix}1\\b^m\\-(\Rho+\nabla_nb^n+b_nb^n)\end{pmatrix}\, .$$}
\be\nn
I^M\!I_M=-\frac{2\sigma^2}{d}\left[\Rho +\nabla^\mu b_\mu -\frac{d-2}{2}\ b^\mu b_\mu\right]\, .
\ee
Since the trace of the $\Rho$ tensor is constant in an Einstein background our model yields a regular scalar mass 
in that case. In fact, since the scale tractor is parallel when the background is conformally Einstein, $I^M I_M$ is then constant for any choice of scale. Moreover, mass is then naturally reintepreted in terms of the Weyl weight of~$\varphi$ 
according to the  mass-Weyl weight relationship
\be
m^2 = -\frac{2\Rho}{d}\Big[\Big(w+\frac{d-1}2\Big)^2-\Big(\frac{d-1}2\Big)^2\Big]\, .\label{massweight}
\ee
Additionally, $\Rho$ is negative and constant in anti de Sitter spaces, therefore reality of  the Weyl weight~$w$ implies the Breitenlohner-Freedman bound~\cite{Breitenlohner:1982jf,Mezincescu:1984ev}  $m^2\geq \frac{\footnotesize\Rho}{2d}\,(d-1)^2$. 

Before studying how spinors couple to scale, it is worth noting how Weyl invariant theories without coupling to the scale appear.
For that, we need the scale $\sigma$ to decouple: examining the Thomas $D$-operator~\eqn{Thomas} at the special weight $w=1-\frac d2$,
we see that its top and middle slots vanish. In that case the equation of motion $I^M D_M \varphi=0$ is equivalent to $D_M\varphi=0$.
This implies $\Big(\Delta - \frac{d-2}{2}\ \Rho\Big)\varphi=0$ which is exactly the equation of motion for a conformally improved scalar field.

\section{Tractor Spinors}
\label{Tractor Spinors}

The theory of spinors in conformal geometry is a well-developed subject to which tractor calulus can be applied~\cite{Helmut,Branson}.
A tractor spinor can be built from a pair of $d$-dimensional spinors. While the latter transform as $\frak{so}(d-1,1)$ representations the tractor spinor
is a spinor representation of $\frak{so}(d,2)$; to avoid technical questions on the spinor type in $d$ and $d+2$ dimensions, we do not specify whether
the constituent $d$-dimensional spinors are Dirac, Weyl, or Majorana. From the  Dirac matrices
$\{\gamma^m,\gamma^n\}=2\eta^{mn}$ in $d$ dimensions,
we build $(d+2)$-dimensional Dirac matrices
\be\nn
\Gamma^+ = \begin{pmatrix}0 & 0\\ \sqrt{2} & 0 \end{pmatrix}, \qquad
\Gamma^- = \begin{pmatrix}0 & \sqrt{2}\\ 0 & 0 \end{pmatrix}, \qquad
\Gamma^m = \begin{pmatrix}\gamma^m & 0\\ 0 & -\gamma^m \end{pmatrix} \, ,
\ee 
subject to 
\be\nn
\{\Gamma^M,\Gamma^N\}=2\eta^{MN}\, .
\ee
A weight~$w$ tractor spinor $\Psi={\tiny\begin{pmatrix}\psi\\\chi\end{pmatrix}}$, built from a pair of $d$-dimensional spinors $\psi$ and $\chi$ is
defined by its transformations under tractor gauge transformations
\be\label{spinorgauge}
\Psi=\begin{pmatrix}\psi\\[2mm]\chi\end{pmatrix}\mapsto \Omega^w \begin{pmatrix}\sqrt{\Omega}\ \psi\\[2mm]\frac1{\sqrt{\Omega}}\Big[\chi
-\frac1{\sqrt{2}} \slashed\Upsilon\psi\Big]\end{pmatrix}\, .
\ee
The tractor covariant derivative acting on a tractor spinor is defined by
\be\nn
\mathcal{D}_\mu \Psi=\begin{pmatrix}   \nabla_\mu \psi + \frac{1}{\sqrt{2}}\gamma_\mu \chi \\[2mm] \nabla_\mu \chi -\frac{1}{\sqrt{2}}\slashed{\Rho}_\mu \psi \end{pmatrix}\, ,
\ee
where $\nabla_\mu$ is the standard Levi-Civita connection acting on $d$-dimensional spinors. 

We have now assembled the ingredients required to compute the Thomas $D$-operator~\eqn{Thomas} acting on spinors. Of particular interest
is the ``Dirac--Thomas $D$-operator''
\bea\nn
\Gamma.D\,   \Psi &=& \begin{pmatrix}(d+2w-2) \slashed\nabla & \frac{1}{\sqrt{2}}(d+2w)(d+2w-2) \\[2mm] -\sqrt{2} \slashed \nabla^2 & -(d+2w)\slashed\nabla   \end{pmatrix}   \Psi \\[4mm]  &=& \begin{pmatrix}(d+2w-2) [ \slashed\nabla \psi + \frac{1}{\sqrt{2}}(d+2w) \chi ] \\[2mm] -(d+2w) \slashed\nabla \chi - \sqrt{2}[\Delta- \frac{\scriptsize \Rho}{2}(d-1)] \psi  \end{pmatrix} \, .\nn
\eea
Here we have denoted the contraction of tractor vector indices by a dot and it is worth bearing in mind that  the $d$-dimensional Weitzenbock
identity acting on spinors is  $\slashed \nabla^2 = \Delta- \frac{\footnotesize \Rho}{2}(d-1)$.

It is well known that the massless Dirac equation is Weyl covariant in any dimension. This follows naturally from tractors:
Observe that we can use the canonical tractor to produce a weight $w+1$ tractor spinor from $\Psi$
\be\nn
\Gamma.X \, \Psi = \begin{pmatrix}0\\[2mm]\sqrt{2} \psi\end{pmatrix}\, ,
\ee
where $\psi$ has the transformation rule
\be
\psi\mapsto \Omega^{w+\frac12}\psi\, .\label{psitr}
\ee
Now acting with the Dirac--Thomas $D$-operator yields 
\be\nn
\Gamma.D \ \Gamma.X \ \Psi = 
(d+2w+2)\ \begin{pmatrix}
(d+2w)\psi\\[2mm]
-\sqrt{2}\slashed \nabla \psi
\end{pmatrix}\, .
\ee
Hence assigning $\Psi$ the weight $w=-\frac d2$ so that $\psi\mapsto \Omega^{\frac{1-d}{2}}\psi$, it follows from the tractor spinor gauge 
transformation rule~\eqn{spinorgauge}, that
\be
\slashed \nabla \psi \mapsto \Omega^{-\frac{d+1}{2}} \slashed\nabla\psi\, ,
\ee
which proves the covariance of the Dirac operator. We are now suitably armed to construct fermionic tractor theories.

\section{Tractor Dirac  Equation}

\label{Tractor Dirac  Equation}

Fermionic theories pose some interesting puzzles for our tractor approach. Firstly, since the tractor approach is based on
arranging fields in $\frak{so}(d,2)$ multiplets, we might generically expect a doubling of degrees of freedom. This can be seen from
the previous Section where tractor spinors were constructed from pairs of spacetime spinors. Secondly, the mass-Weyl weight
relationship~\eqn{massweight} relates the mass squared to the scalar curvature. However, massive spinor theories depend
linearly on the mass, and therefore the square root of the scalar curvature. It is not immediately obvious how this square root could arise.
We will solve both of these puzzles by employing several principles: To construct tractor-spinor and tractor-spinor-vector theories
\begin{enumerate}
\item We search for massive wave equations whose masses are related to Weyl weights by an analog of the scalar relationship~\eqn{massweight}.
\item We require that, in a canonical choice of scale, these theories match those found by the log-radial dimensional reduction of
$d+1$ dimensional massless Minkowski theories to $d$ dimensional constant curvature ones described in~\cite{Biswas:2002nk,Hallowell:2005np}
and Appendix~\ref{Doubled Reduction}.
\item We will impose as many constraints as consistent with the above requirements so as to find a ``minimal covariant field content''.
\end{enumerate}
These principles will become clearer through their applications, so let us provide the details.

In our previous work, we used the fact that (from an ambient viewpoint as described in~\cite{CG,GP} and further studied in~\cite{Gover:2009vc}),
 the contraction of the
scale tractor and Thomas $D$-operator $I.D$ generates bulk evolution, and therefore searched for wave equations $(I.D+\mbox{more})V^\bullet=0$,
where the terms ``+more'' were chosen on the grounds of gauge invariance. In addition we imposed the most general  field constraints, linear in the Thomas $D$-operator, on the bosonic gauge fields $V^\bullet$ that were consistent with gauge invariance. In this picture the tractor  weight $w$ of the gauge field $V^\bullet$, controls the mass, save at special weights where the theory becomes massless or partially massless. In addition at the special weight $w=1-d/2$ the scale tractor
decouples from the equations of motion and conformal wave equations result (a comprehensive study of higher spin conformal wave equations may be found in~\cite{Vasiliev:2009ck,Bekaert:2009fg}, see also references therein).

Along the same lines we propose the spinor equation of motion and field constraint for a weight~$w$ tractor spinor~$\Psi$
\bea\nn
I.D\ \Psi&=&0\, ,\nn\\[2mm]
\Gamma.D\ \Psi&=&0\, .\label{dirac1}
\eea
We can view the second equation as a scale covariant constraint eliminating the lower component of~$\Psi$.
Its solution is
\be\nn
\Psi=
\begin{pmatrix}
\psi\\[2mm]
-\frac{\sqrt{2}}{d+2w}\, \slashed{\nabla} \psi\
\end{pmatrix}\, .
\ee
In turn, the $I.D$ field equation, in the canonical scale $\sigma=$ constant, implies the massive wave equation\footnote{The value $w=-d/2$ is distinguished here, as in  fact is the value $w=-d/2+2$. In the first case we cannot solve the constraint in~\eqn{dirac1}. Also, in deriving~\eqn{wave}, we have dropped 
an overall factor $(d+2w-2)/(d+2w)$. However, below we give a second formulation of the system that still predicts~\eqn{wave} at $w=-d/2+2$.} 
\be\label{wave}
\Big[\Delta + \frac{2\Rho}{d}(w^2+wd+\frac d4)\Big]\,  \psi=0\, .
\ee
Defining the squared mass as the eigenvalue of $\Delta$ (note that  $\Rho$
is constant in an Einstein background) gives the spinorial mass-Weyl weight relationship
\be\label{massweyl}
m^2= -\frac{2\Rho}{d}\Big[\Big(w_\psi+\frac{d-1}2\Big)^2-\frac{d(d-1)}{4}\Big]\, ,
\ee
analogous to its bosonic counterpart~\eqn{massweight}. Here we have defined $w_\psi\equiv w+\frac12$ because
under Weyl transformations $\psi$ transforms according to~\eqn{psitr}.
Observe that reality of the weight $w$ for spaces with negative scalar curvature implies a Breitenlohner--Freedman type bound~\cite{Breitenlohner:1982jf, Mezincescu:1984ev} on the mass parameter  $m^2\geq\frac12\Rho(d-1)$. 
Before analyzing this system further, let us present an alternate formulation.

The Thomas $D$-operator is second order in its lowest slot. For Fermi systems, we would like to find a set of
first order field equations. To that end, we recall the double $D$-operator defined by
\be\label{DDXD}
(d+2w-2)D^{MN}=X^N D^M - X^M D^N\, .
\ee
At generic weights it obeys the identity
\be\nn
[X^M,D^N]=2 D^{MN} - (d+2w)\, \eta^{MN}\, ,
\ee
so the double-$D$ operator essentially amounts to the commutator of the Thomas $D$-operator and the canonical tractor.
(An ambient interpretation of this algebra is explored in~\cite{Gover:2009vc}.)
In components it is given by
\be\label{DD}
D^{MN}=
\begin{pmatrix}
0&0&w\\[1mm]
0&0&{\cal D}^m\\[1mm]
-w&-{\cal D}^n&0
\end{pmatrix}\, .
\ee
In these terms, we propose the Dirac-type equation\footnote{This equation is similar in spirit to Dirac's proposal for writing four dimensional conformal wave equations by employing the six dimensional Lorentz generators~\cite{Dirac:1936fq}. Of course here, we also describe massive systems that are not
invariant without coupling to scale.}
\be\label{dirac2}
I^M\Gamma^N D_{MN}\, \Psi=0\, .
\ee
In the canonical choice of scale it reads
\be\nn
-\sigma\begin{pmatrix}\slashed\nabla\psi +\frac{d+2w}{\sqrt{2}}\, \chi\\[2mm]-\slashed\nabla\chi+\frac{(d+2w)\!\!{\scalebox{.7}{ \Rho}}}{\sqrt{2}\, d}\, \psi\end{pmatrix}=0\, .
\ee

Firstly, when $d+2w\neq0,2$, it is easy to verify that these equations are equivalent to~\eqn{dirac1}. In general, they are more fundamental
because (at $d+2w\neq2$) the equation~\eqn{wave} follows as an integrability condition.
Moreover, even 
 at $d+2w=2$ we can still define the double-$D$ operator by~\eqn{DD} and then have well-defined system (that implies the massive wave equation~\eqn{wave}).

The weight $d+2w=0$, has a special physical significance because at that weight we expect to find a scale invariant theory as explained in Section~\ref{Tractor Spinors}. Although it is not true that the scale decouples from the equation~\eqn{dirac2}, the modified equation
$$
\sigma^{-1}\Gamma.X I^M\Gamma^N D_{MN}\, \Psi=0\, ,
$$
is in fact independent of the scale at $w=-d/2$. It is then equivalent to the equation $\Gamma.X\  \Gamma^N D_{MN}\Psi=0$
(just as for scalars in Section~\ref{Coupling to Scale}). In components this amounts simply to the Dirac equation $\slashed\nabla\psi=0$.

In the above formulation, $w=-d/2$ is the only value at which multiplication by a factor $\Gamma.X$ yields a consistent system, at other values
the field~$\chi$ enters on the right hand side of the Dirac equation.
The presence of the second spinor~$\chi$ is undesirable, because it doubles the degrees of freedom of the $d$-dimensional theory.
We next explain how to obtain a tractor theory of a single $d$-dimensional spinor.   

Firstly observe that a massive Dirac equation is linear in the mass parameter, whereas according to~\eqn{massweyl} the constant scalar
curvature is proportional to the square of the mass. Therefore we need a tractor mechanism that somehow introduces the square root of
the scalar curvature while at the same time relating the pair of spinors  $\chi$ and $\psi$. Examining our spinorial wave equations~\eqn{dirac2}
in their canonical component form, we see that a relationship $\chi=\alpha\psi$ means that this pair of equations are equivalent only when $\alpha^2=-\Rho/d$. This relationship
can be imposed tractorially using the projectors
\be\nn
\Pi_\pm \equiv \frac12\Big[ 1\pm \frac{\Gamma.I}{\sqrt{I.I}}\Big]\, .
\ee
(Recall that in a conformally Einstein background, $I^M$ is tractor parallel, so that $I.I$ is constant. Note that $I.I$ is positive
for negative scalar curvature.)
Hence we propose the tractor Dirac equations
\be\nn
I^M \Gamma^N D_{MN} \Psi = 0 = \Pi_+\Psi\, ,
\ee
We could equally well multiply the first of these equations by $\Gamma.X$ since the its bottom slot is a consequence of the top one.
(The choice of $\Pi_+$ rather than $\Pi_-$ corresponds to the sign of the Dirac mass term.) In canonical components these
imply the massive curved space Dirac equation
\be\label{massdirac}
\Big[\slashed\nabla -\sqrt{\frac{-\Rho}{2d}}\ (d+2w) \Big]\psi=0\, .
\ee
Its  mass is again related to the weight of $\psi$ by~\eqn{massweyl}.

In summary, the irreducible tractor Dirac equation for a tractor spinor $\Psi$ (subject to the ``Weyl''-like condition\footnote{We cannot help but remark that this conditions melds two of Weyl's seminal contributions to physics -- the Weyl spinor and Weyl symmetry.} $\Pi_+ \Psi=0$) is given by
$$
\Gamma.X I^M \Gamma^N D_{MN} \Psi=0\, .
$$ 
This equation of motion follows from an action principle which we now describe.
To that end we need to 
introduce the tractor Dirac conjugate spinor, which is defined as
\be\nn
\overline\Psi \equiv \overline{\!\begin{pmatrix}\psi\\\chi\end{pmatrix}\!}=i \Psi^{\dagger} \Gamma^{\bar 0}=(\bar\chi \;\; \bar\psi)\, ,
\ee
where $\bar\psi$ and $\bar\chi$ are the standard $d$-dimensional Dirac conjugates of $\psi$ and $\chi$, and $ \Gamma^{\bar 0}$ obeys the following properties (because it derives from the profuct of the two timelike Dirac matrices of $\frak{so}(d,2)$):
\be \nn
(\Gamma^{\bar 0})^{2} = -1 \, , \qquad \Gamma^{\bar 0 \dagger} =- \Gamma^{\bar 0} \, , \qquad \Gamma^{M \dagger} = - \Gamma^{\bar 0} \Gamma^M \Gamma^{\bar 0} \, .
\ee
Then the required action principle is
\be
S=\frac 1{\sqrt{2}}\int \frac{\sqrt{-g}}{\sigma^{d+2w+1}}\overline\Psi \Gamma.X I^M \Gamma^N D_{MN} \Psi\, ,
\ee
where the tractor spinor $\Psi$ obeys $\Pi_+\Psi=0$. This action is hermitean. Since it is useful to possess the tractor machinery
required to vary actions of this type, let us prove this. Firstly, the double-$D$ operator $D_{MN}$ is Leibnitzian. 
Moreover $\int \sqrt{-g} D_{MN} \Xi^{MN} = 0$ (up to surface terms) for any $\Xi^{MN}$ of weight zero. This allows us to integrate $D_{MN}$ 
by parts. Therefore,  to verify $S=S^\dagger$ we need to compute $D_{MN} \Big[\frac{1}{\sigma^{d+2w+1}}I^M \Gamma^N \Gamma.X \Psi\Big]$
which requires the following identities
\bea
D_{MN}\sigma\ \ \ &=&X_N I_M-X_M I_N\, ,\nn\\[2mm]
D_{MN}I^N\  &=&0\, ,\nn\\[2mm]
D_{MN} X^R \ &=&X_N \delta_M^R - X_M \delta_N^R\, ,\nn\\[2mm]
X^MD_{MN} &=&w X_N\, .
\eea
Orchestrating these, we find
\be\nn
S-S^\dagger=\frac{d+2w}{\sqrt{2}}\int \frac{\sqrt{-g}}{\sigma^{d+2w+1}}\ \overline\Psi (\Gamma.X \Gamma.I -\sigma)\Psi\, .
\ee

%%NEEDS ATTENTION!!
For generic weights $w$ this is non-vanishing, however using the condition $\Pi_+\Psi=0$ to conclude that $\Psi$ is in the image of $\Pi_-$
along with the facts that $\overline {\Pi_- \Psi}\equiv \overline\Psi \Pi_-$ and $\Pi_-(\Gamma.X \, \Gamma.I -\sigma)\Pi_-=0$, shows that $S=S^\dagger$. 
%(in fact any projective condition $\frac12 (1\pm \frac{V.\Gamma}{\sqrt{V.V}})\Psi=0$ for some vector $V^M$ on $\Psi$ would suffice). 
A similar  computation 
implies that the above action implies the field equations quoted.
%%%

Our final computation is to  write out the action principle in components. Rather than working at the canonical scale, lets us give the general result,
namely
\be
\nn
S=-\int\frac{\sqrt{-g}}{\sigma^{d+2w}}\ \bar\psi \left[\slashed\nabla - \frac12(d+2w)\, \Big(\slashed b +\sqrt{-\frac{2(\Rho+\nabla\cdot b -\frac{d-2}{2}\ b\cdot b)}{d}}\:  \Big)\right]\ \psi\, .
\ee
Each term has a simple interpretation.  The $\slashed b$ contribution covariantizes the leading Dirac operator with respect to scale transformations
so,   $\psi\mapsto \Omega^{w+\frac 12}\psi$ implies $[\slashed \nabla - \frac12 (d+2w) \slashed b] \psi \mapsto \Omega^{w+\frac12} [\slashed \nabla - \frac12 (d+2w) \slashed b]\psi$. These terms also follow from the standard Weyl compensator mechanism. The square root factor is the mass term which
equals (up to a factor $\sigma$) $\sqrt{I.I}$, and is therefore constant for conformally Einstein backgrounds.  The prefactor $(d+2w)$ calibrates the mass
to the square of the scale tractor and implies the mass-Weyl weight relationship~\eqn{massweyl}. When $w=-\frac d2$, the scale $\sigma$ decouples from
the action and we obtain the Weyl invariant curved space Dirac equation discussed in Section~\ref{Tractor Spinors}.

\section{Tractor Rarita--Schwinger Equation}

\label{Tractor Rarita--Schwinger Equation}

In the spinor models we have encountered so far there have been choices for  mass terms: we could have used a ``gravitational mass term''~\eqn{grav_mass_term} or a compensated mass term.  The gravitational mass term is proportional to $I.I$, while the compensated mass term is obtained by using the scale $\sigma$ to compensate a standard mass term (for example,
we could add a term $\frac12\int \frac{\sqrt{-g}}{\sigma^{d+2w}}\  \varphi^2$ to the scalar action principle). However, once we study models
with spins $s\geq1$, gauge invariances are necessary to ensure that only unitary degrees of freedom propagate. In our previous work~\cite{Gover:2008sw, Gover:2008pt}, we showed
how higher spin gauge invariant tractor models described bosonic massless, partially massless~\cite{Deser:1983tm,Higuchi:1986py,pm} and massive models in a single framework.
In particular, they implied  ``gravitational mass terms'' (rather than compensated ones) with masses dictated by Weyl weights. We now extend those 
results to the higher spin $s=3/2$ Rarita--Schwinger system. The following analysis closely mirrors the tractor Maxwell system studied in~\cite{Gover:2008sw, Gover:2008pt}
so we keep details to a minimum.

As field content, we take a weight $w$ tractor vector-spinor $\Psi^M$ subject to the gauge invariance
\be\nn
\delta \Psi^M = D^M \Xi\, ,
\ee
where $\Xi$ is a weight $w+1$ tractor spinor parameter. Since the Thomas $D$-operator is null, we may consistently
impose the field constraint 
\be\label{field}
D_M \Psi^M=0\, .
\ee
 We assume that the background is conformally flat, so that Thomas $D$-operators commute\footnote{We leave an investigation
of whether non-minimal couplings could relax this restriction to future work. Any such study will be highly constrained by existing
results for gravitational spin~3/2 couplings, see~\cite{Deser:2001dt}.}. We now observe that the quantity
\be\nn
{\cal R}^{MNR}=3D^{[MN}\Psi^{R]}\, ,
\ee
is gauge invariant by virtue of the identity~\eqn{DDXD} and use it to construct a set of tractor Rarita--Schwinger equations
coupled to the scale tractor
\be\label{rarita}
{\cal R}_M\equiv\Gamma_{MNR} I_S {\cal R}^{SNR}=0\, .
\ee
The final requirement we impose is the projective one found for spinors
\be\label{project}
\Pi_+\Psi^M=0\, , \qquad  \Pi _{-} \Xi = 0.
\ee
To verify that the set of equations~(\ref{field},\ref{rarita},\ref{project})
are the desired ones, we write them out explicitly in canonical components. This computation is lengthy but straightforward.
The field constraint~\eqn{field} and projective condition~\eqn{project} eliminate most of the field content leaving only
the top spinorial components~$\psi^+$, and middle vector slots~$\psi^m$, independent. Since the system will describe both massive and massless excitations, the spinor $\psi^+$ plays the {\it r\^ole} of a St\"uckelberg field.
The Rarita--Schwinger type equation ${\cal R}^M=0$ in~\eqn{rarita} then yields the independent field equations
\be\nn
\gamma^{\mu\nu\rho}\wt \nabla_\nu \psi_\rho +\sqrt{-\frac{2\Rho}{d}}\ \gamma^{\mu\nu} \Big([w+1]\psi_\nu-  \wt \nabla_\nu \psi^+\Big)=0\, .
\ee
Here the operator 
\be\nn
\wt \nabla_\mu = \nabla_\mu - \sqrt{\frac{-\Rho}{2d}}\gamma_\mu
\ee
is the modification of the covariant derivative acting on spinors found quite some time ago in a cosmological supergravity
context~\cite{Townsend:1977qa}. Its distinguishing property is that $[\wt\nabla_\mu,\wt \nabla_\nu]$ vanishes on spinors (but not vector-spinors).
The above equation of motion enjoys the gauge invariance
\bea\nn
\delta\psi_\mu &=& (d+2w) \wt \nabla_\mu \varepsilon\, ,\nn\\[2mm]
\delta\psi^+ &=& (d+2w) (w+1) \varepsilon\, .
\eea
We include the factor $(d+2w)$ to synchronize the component transformations  with the tractor ones $\delta \Psi^M=D^M\Xi$.
Notice  they imply that $\psi^+$ is an auxiliary St\"uckelberg field at generic $w\neq 1$, which can be gauged away 
leaving a massive Rarita--Schwinger field $\psi_\mu$. When $w$ does equal~$-1$, the field $\psi^+$ is gauge inert
and we may impose the additional constraint $\psi^+=0$ (in fact, a careful analysis shows that this field decouples completely
at $w=-1$). That leaves the massless Rarita--Schwinger equation  in AdS with standard gauge invariance
\be\nn
\gamma^{\mu\nu\rho}\wt \nabla_\nu \psi_\rho=0\, ,\qquad \delta \psi_\mu = \wt \nabla_\mu \varepsilon\, .
\ee
Returning to generic $w$, we may rewrite the above equation in the standard massive form
\be\nn
\gamma^{\mu\nu\rho}\nabla_\nu\psi_\rho + m\gamma^{\mu\nu}\psi_\nu = 0\, .
\ee
The integrability conditions for this system imply the usual constraints $\nabla^\mu\psi_\mu$ $=$ $0$ $=$ $\gamma^\mu\psi_\mu$,
and in turn $(\slashed \nabla - m)\psi_\mu=0$. The mass $m$ is here given in terms of weights by the mass-Weyl weight
relationship
\be\nn
m=\sqrt{-\frac{\Rho}{2d}}\ (d+2w)\, .
\ee
Via the spin~3/2 Weitzenbock identity, this implies a wave equation $(\Delta-\mu^2)\psi_\mu=0$
where $\mu^2$ obeys a Weyl weight relationship highly reminiscent of the spin~0 and~1/2 ones above
\be \label{m32}
\mu^2 = -\frac{2\Rho}{d}\Big[\Big(w_{\psi_m}+\frac{d-1}{2}\Big)^2-\frac{d(d-1)}{4}-1\Big]\, .
\ee 	
Here $w_{\psi_m}=w+\frac12$ because, in the St\"uckelberg gauge $X.\Psi=0$, we have
the Weyl transformation rule $\psi_\mu\mapsto \Omega^{w+3/2}\psi_\mu$. We end by observing,
that this result implies a Breitenlohner--Freedman type bound for massive gravitini $\mu^2\geq \frac{\!\scalebox{.7}{\Rho}}{2d}\ [d(d-1)+4]$.
Following the procedure outlined at the end of Section~\ref{Tractor Dirac  Equation}, we write~\eqn{rarita} at arbitrary scale
\be
\nn
\gamma^{\mu\nu\rho}\nabla_\nu\psi_\rho + \frac{1}{2}(d+2w) \gamma^{\mu\nu}\Big(\slashed b + \frac{\sqrt{I^2}}{\sigma}\Big)\psi_\nu - (w+1)\gamma^{\mu} b \cdot \psi= 0\,\, .
\ee
To understand the above expression, let us define the Weyl-covariantized Rarita-Schwinger operator
 \be
R^{\mu} \equiv  \gamma^{\mu\nu\rho}\nabla_\nu\psi_\rho + \frac{1}{2}(d+2w) \gamma^{\mu\nu}\slashed b \psi_\nu - (w+1)\gamma^{\mu} b \cdot \psi  \,  \, , \qquad \gamma \cdot \psi =0 \, .
\ee
The $b$ contribution covariantizes the Rarita-Schwinger operator with respect to scale transformations
such that $\psi_\mu \mapsto \Omega^{w+\frac32} \psi_\mu$ implies  $R^\mu\mapsto \Omega^{w-\frac32} R^{\mu}$, 
modulo the condition $\gamma\cdot\psi=0$. This operator also follows from the standard Weyl compensator mechanism. As before, the square root factor is the mass term,
and the  prefactor $(d+2w)$ calibrates the mass to the square of the scale tractor and implies the mass-Weyl weight relationship~\eqn{m32}.  In $d=2$, when $w=-\frac d2=-1$, the scale $\sigma$ decouples from
the equation of motion and we obtain the Weyl invariant curved space Rarita-Schwinger equation.

In fact, in arbitrary dimensions $d$ it is possible to write down a Weyl invariant Rarita--Schwinger system~\cite{Deser:1983tm}.
We can obtain that theory from our tractor one as follows:
Consider  a new field equation $\tilde R^{\mu} = R^{\mu} - \frac{1}{d} \gamma^{\mu} (\gamma \cdot R)$ $=$ $0$, or explicitly
\be
\tilde R^{\mu} = \slashed \nabla \psi^{\mu} -\frac{2}{d} \gamma^{\mu} \nabla \cdot \psi + \frac{d+2w}{2}[ \gamma^{\mu\nu}\slashed b \psi_\nu - \frac{2(d-1)}{d} \gamma^{\mu} b \cdot \psi]\, .
 \ee
When $w= -\frac{d}{2}$, the scale dependence through the composite gauge field $b$ decouple completely, and we are left with the Weyl invariant Rarita-Schwinger system of~\cite{Deser:1983tm} generalized to arbitrary dimensions
\be \label{weyl32}
\slashed \nabla \psi^{\mu} -\frac{2}{d} \gamma^{\mu} \nabla \cdot \psi = 0 = \gamma\cdot \psi\, .
\ee
We can derive the same results efficiently using  tractors.  This requires  imposing two additional constraints
\be
X . \Psi = 0 \, ,\qquad \Gamma . X\  \Gamma . \Psi = 0 \,,
\ee
which in components read 
\be  
\psi^+ = \chi^+=\gamma \cdot \psi = 0.
\ee
As argued before, at $ w= \frac{-d}{2}$ the compensator field $\sigma $ can be safely eliminated without compromising the Weyl invariance.  At this special value of the weight, the tractorial expression describing Weyl invariant Rarita-Schwinger equation is\footnote{Note that there actually no pole in this expression in six dimensions as evidenced by the component expression~\eqn{weyl32}} 
\be
\tilde R^M = \Gamma . X [R^M - \frac{d-2}{d(d-6)}\Gamma^M (\Gamma . R)]=0 \, ,
\ee 
which in components exactly matches~\eqn{weyl32}.

\section{Supersymmetry}

\label{Supersymmetry}

Given a tractor description of spinors and scalars, it is natural to search for 
a supersymmetric combination of the two. Here we study global supersymmetry.
In a curved background, globally supersymmetric theories require a generalization
of the constant spinors employed as parameters of supersymmetry transformations
in flat space. A possible requirement is to search for covariantly constant spinors,
although most backgrounds do not admit such special objects. Focusing on 
conformally flat backgrounds, a more natural condition is to require that the background
possess a Killing spinor $\varepsilon$ defined by
\be\nn
\nabla_\mu\varepsilon = -\sqrt{\frac{-\Rho}{2d}}\ \gamma_\mu\varepsilon\, .
\ee
As a consequence it follows that $\bar\varepsilon\varepsilon$ is constant.
This condition can be neatly expressed in tractors in terms of what we shall call a ``scale spinor''
\be\nn
\Xi=\begin{pmatrix}\varepsilon\\[1mm]\eta\end{pmatrix}\, ,\qquad \Pi_-\Xi=0\, .
\ee
Here $\eta$ is determined by the projective condition. The Killing spinor condition for $\varepsilon$ is
now imposed by requiring the weight $w=0$ tractor spinor $\Xi$ to be tractor parallel
\be\nn
{\cal D}_\mu\Xi=0\, .
\ee
From the scale spinor, we can form the scale tractor as
\be\nn
I^M=\frac{\sigma\, \overline\Xi\Gamma^M\Xi}{\, \overline\Xi \Gamma.X \Xi}\, ,
\ee
which justifies its name.

Having settled upon the global supersymmetry parameters, we specify the field content as a weight $w+1$ scalar~$\varphi$
and a weight $w$ tractor spinor~$\Psi$  subject to
\be\nn
\Pi_+\Psi=0\, .
\ee
We have chosen the tuning between weights of fermionic and bosonic fields in order to preserve supersymmetry.
The supersymmetry transformations are given by\footnote{There is no pole in the fermionic variation at $w=-d/2$; this can be checked explicitly from a component computation.}
\bea
\delta\varphi &=& \Re \Big(  \overline\Xi \Gamma.X \Psi\Big)\, ,\label{susyb} \\ 
\delta \Psi &=& \frac{1}{d+2w} \Big[(\Gamma.D-\frac1\sigma\,  \Gamma.X I.D)\varphi\Big]\, \Xi\, .\label{susyf}
\eea
Here we take $\varphi$ to be real, but make no assumption for reality conditions for the spinors. If the underlying $d$-dimensional
spinors are Majorana, there is no need to take the real part in the supersymmetry transformations of the bosons. For the independent
bosonic and fermionic field components, these transformations amount to
\be\nn
\delta\varphi=\Re(\sqrt2\bar\varepsilon\psi)\, ,\qquad\delta\psi=\Big[\Big(\slashed\nabla+\sqrt{\frac{-2\Rho}{d}}\ (w+1)\Big)\varphi\Big]\varepsilon\, .
\ee
The invariant tractor action for this system is the sum of the Bose and Fermi actions discussed in previous Sections
\be\nn
S=\int \frac{\sqrt{-g}}{\sigma^{d+2w+1}}\, \Big\{\overline\Psi \Gamma.X \, \Gamma^M I^N D_{MN}\Psi + \varphi I.D \varphi\Big\}\, .
\ee
To verify the invariance of this action one first uses the identity
\be\nn
\Gamma.X\,\Gamma^MI^N D_{MN}=\frac{\sigma}{d+2w-2} \Gamma.X\,  \Gamma.D\, ,
\ee
so that
\be\nn
\Gamma.X\  \Gamma^MI^N D_{MN}\delta\Psi=-\Big(\frac\sigma{(d+2w-2)(d+2w)}\Gamma.X \ \Gamma.D \ \Gamma.X \ \frac1\sigma\  I.D\varphi\Big)\Xi\, .
\ee
Then the identity
\be\nn
\Gamma.X\ \Gamma.D=-\frac{d+2w-2}{d+2w+2} \Gamma.D\ \Gamma.X +(d+2w)(d+2w-2)\, ,
\ee
yields
\be\nn
\overline\Psi \Gamma.X\  \Gamma^MI^N D_{MN}\delta\Psi=-(\overline\Psi \Gamma.X\Xi) \ I.D\varphi.
\ee
Comparing the last expression with the bosonic variation in~\eqn{susyb} completes our invariance proof.

\section{Interactions}

\label{Interactions}

To add interactions, we begin by closing the supersymmetry algebra off-shell with the aid of an auxiliary field.
In curved backgrounds, the square of a supersymmetry transformation yields an isometry as the generalization of
translations in flat space. Therefore we also need to explain how to handle isometries with tractors. On the bosonic field $\varphi$,
the supersymmetry algebra closes without any auxiliary field and the algebra of two supersymmetry transformations is given by
\be\nn
[\delta_1,\delta_2]\varphi = \Re(\overline\Xi_1\Gamma^{MN}\Xi_2) \, D_{MN} \varphi\, .
\ee
The adjoint tractor $\Re(\overline\Xi_1\Gamma^{MN}\Xi_2)$ is an example of what we shall call a ``Killing tractor''~\cite{KT,GP}. Let us make a brief aside to 
describe these objects:
Suppose that $\xi^\mu$ is any vector field. Then we can form a weight $w=1$ tractor
\be\nn
V^M=\begin{pmatrix}0\\[1mm] \xi^m\\[2mm]-\frac1d \nabla_\mu \xi^\mu\end{pmatrix}
\ee
subject to $X.V=D.V=0$. In turn we may build an adjoint tractor
\be\nn
V^{MN}=\frac1{d} D^{[M} V^{N]} =
\begin{pmatrix}
0 & \xi^n & -\frac 1d \nabla_\mu\xi^\mu\\[1mm]
\mbox{a/s} & \nabla^{[m}\xi^{n]}& 
%\frac1{2d}\Big[\nabla^\mu\Big(\nabla_\mu \xi^m +\nabla^m \xi_\mu -\frac2d e_\mu{}^m\nabla^\nu\xi_\nu\Big)-4\Rho^m_\mu \xi^\mu -2 \nabla^m \nabla^\mu\xi_\mu\Big]
\frac{1}{2d}\Big([\Delta+\Rho]\xi^m \!-\!\frac{d+2}{d}[\nabla^m \nabla_\mu +d\Rho^m_\mu]\xi^\m\Big)
\\[3mm]
\mbox{a/s}&\mbox{a/s}&0
\end{pmatrix} \, .
\ee 
The operator 
\be\nn
%{\cal L}_\xi\equiv 
\frac12 V^{MN} D_{NM} = \xi^\mu{\cal D}_\mu -\frac wd (\nabla_\mu\xi^\mu)\, ,
\ee
may be viewed as a tractor analog of the vector field $\xi^\mu\partial_\mu$.
Notice that acting on weight $w$ scalars, it gives the correct transformation law
for a conformal isometry
\be\nn
\delta \varphi = (\xi^\mu \partial_\mu  -\frac wd[\nabla_\mu\xi^\mu])\varphi\, .
\ee
It is not difficult to verify that $\Re(\overline\Xi_1\Gamma^{MN}\Xi_2)$ corresponds to $V^{MN}$ with
$\xi^\mu$ given by the Killing vector $\sqrt{2}\Re(\bar\varepsilon_1\gamma^\mu\varepsilon_2)$.
This shows that acting on $\varphi$, the supersymmetry algebra closes onto  
 isometries\footnote{It could be interesting and natural in our framework to study extensions where the supersymmetry algebra closes onto conformal isometries.}. 

To close the algebra on the fermions we need first to understand how (conformal) isometries act on (tractor) spinors.
In the work~\cite{Gover:2009vc}, the double $D$-operator was related to the generators of ambient Lorentz transformations. This suggests
that, acting on tractors of arbitrary tensor type, we should  introduce the operator
\be\nn
\pounds = \frac 12 V^{MN}\Big[D_{NM}+ {\cal S}_{MN}\Big]\, ,
\ee
where ${\cal S}_{MN}$ are the ambient intrinsic spin generators. On spinors we have
\be\nn
{\cal S}_{MN}=\frac12 \Gamma_{MN}\, .
\ee
Indeed, the transformation rule $\delta\Psi = \pounds \Psi$ for a weight~$w$ tractor spinor $\Psi$ with top slot $\psi$ implies
\be\nn
\delta\psi = (\pounds_\xi  - \frac{w_\psi}{d}[\nabla_\mu\xi^\mu])\psi\, ,
\ee
where the Lie derivative on spinors is $\pounds_\xi\psi = (\xi^\mu\nabla_\mu + \frac14 \gamma^{\mu\nu}[\nabla_\mu\xi_\nu])\psi$.

To obtain a closed, offshell supersymmetry algebra for the fermions we need to introduce auxiliary fields. Since the off-shell bosonic and fermionic
field contents must balance, the details will depend on the dimensionality. Therefore, for simplicity, we now restrict ourselves to a four dimensional 
chiral multiplet with $(z,\psi,F)$ where $z$ and $F$ are complex scalars and $\psi$ is a Majorana spinor. We represent the scalars $z$ and $F$ by  
weight $w+1$ and $w$ tractor scalars with the same names while $\psi$ is the top slot of a weight $w$ tractor spinor $\Psi$  subject to $\Pi_+\Psi=0$.
Notice, this implies that independent spinor field content is characterized by 
\be\nn
\Gamma.X \Psi = \sqrt{2}\, \begin{pmatrix}0\\ \psi\end{pmatrix}\, .
\ee
It is therefore sufficient (and simplifying) to specify the transformation rules of $\Gamma.X\Psi$ in what follows
(note also that the operators $\Gamma.X$ and $\pounds$ commute). 
The supersymnmetry transformations of our tractor chiral multiplet then read
\bea
\delta z\quad\  &=& \overline\Xi \Gamma.X {\cal L} \Psi\, ,\nn\\
\delta (\Gamma.X {\cal L}\Psi) &=&\Gamma.X {\cal L} \Big( F + \frac{1}{d+2w}\Gamma.D z\Big)\Xi\, ,\nn\\
\delta F\quad &=& \frac{-1}{d+2w+2}
\, \overline\Xi \Gamma.D \Gamma.X {\cal L}\Psi\, . 
\eea
The rules for the complex conjugates are given by replacing ${\cal L}\mapsto{\cal R}$ where ${\cal L,R}=\frac12 (1\mp \Gamma^7)$, or explicitly
\be\nn
{\cal L,R}=\begin{pmatrix}L,R & \\ & R,L\end{pmatrix}\, ,\qquad L,R = \frac12 (1\mp \gamma^5)\, .
\ee
In components, these transformation rules agree with the usual ones for a massive Wess--Zumino model in an AdS background~\cite{Ivanov:1979ft,Dusedau:1985uf, Burges:1985qq}.
It is important to note however, since the tractor system treats the massive and massless systems on the same footing, the auxiliary field used here
differs from the standard one by terms linear in the complex scalar $z$.

Closure of the supersymmetry algebra on the scalars can be verified as described above. For the Fermions, a tractor Fierz identity is required
\be\nn
{\cal R}\, \Xi_2\overline\Xi_1{\cal R}-(1\leftrightarrow2) = -\frac 18 (\overline\Xi_1\Gamma^{RS}\Xi_2) \ {\cal R}\, \Gamma_{RS}{\cal R} \, .
\ee
After some algebra it follows that
\be\nn
[\delta_1,\delta_2](\Gamma.X {\cal L} \Psi) = \pounds \, \Gamma.X {\cal  L} \Psi\, , \qquad [\delta_1,\delta_2] F = \pounds F\, ,
\ee
proving that the supersymmetry algebra closes. 

Armed with a closed supersymmetry algebra, an invariant action principle is easily obtained in  tractors:
\be
S = \int  \frac{\sqrt{-g}}{\sigma^{d+2w}}(\mathcal{L}_{\rm kin} + \mathcal{L}_{\rm int}) \, ,
\ee
where\\[2mm]
\bea
\mathcal{L}_{\rm kin} &=&  \frac{1}{2\sigma}\bar\Psi\Gamma. XI_M \Gamma_N D^{MN} \Psi - \frac{1}{\sigma}\bar z I . D z + | F|^2  \nn\\
&-& 2a(w+2) \big[(F - \bar F)(z- \bar z) +  a(2w+5)(z -\bar z)^2  \big]  \, , \\[2mm]
\mathcal{L}_{\rm int}  &=& \frac{1}{2}\bar\Psi \Gamma. X(\mathcal{L}W'' + \mathcal{R}\bar W'') \Psi\nn\\
&-&   FW' - \bar F \bar W' - 2a (w+1)(zW' + \bar z \bar W') -6a (W + \bar W)  \, .\nn\\
\eea
Notice the appearance of the weight $-1$ scalar $a$ in the weight  $2w$ lagrangian density.  At arbitrary scales $$ a =\frac{\sqrt{I.I}}{2\sigma}\, ,$$
while in a canonical choice of scale it is related to the four dimensional cosmological constant by $12a^2=-\Lambda$.

The action is split into a kinetic  and interacting pieces. The latter depends linearly on a weight $2w+1$ holomorphic potential~$W=W(z,\sigma,a)$. Since our tractor theories describe massive and massless excitations uniformly in terms of weights, the above splitting is not the canonical
one into a massless action plus potential terms, but rather uses the freedom of the function~$W$ to split the action into  free (generically massive) and interacting pieces. At $w=-2$. the Lagrangian density ${\cal L}_{\rm kin}$, expressed in components at the canonical choice of scale recovers the massless part of the supersymmetric AdS Wess--Zumino model quoted in~\cite{Dusedau:1985uf,Burges:1985qq}.

%With the rescaling, $ W = \sigma a \tilde W $,  $ F = \sigma a \tilde{F}$, and $ \Xi = \sigma a \bar \Xi$, the lagrangian takes a simpler form:
%\bea
%\mathcal{L} &=&  I^2 \mathcal{F} + \frac{1}{2}\bar\Psi\Gamma. XI_M \Gamma_N D^{MN} \Psi +(d+2w) \bar z I . D z + \frac{1}{4}\bar\Psi \Gamma. X(\mathcal{L}W'' + \mathcal{R}\bar W'') \Gamma. I\Psi\, ,
%\eea
%where
%\bea
%\mathcal{F} &=& \frac{(d+2w)}{4} \{ FW' + \bar F \bar W' + 2\sigma^{-1} [(w+1)(zW' + \bar z \bar W') + (d-1)(W + \bar W) ] +  \sqrt{2}(F \bar F \nn  \\[2mm]
%&+&   \sigma^{-1}(d+2w)(z \bar F + \bar z F) + 2\sigma^{-2}(d+2w)(d+2w+1) z \bar z ) \} \, .
%\eea
%\edz{This is a mess, how do they know what a is??}

\section{Conclusions}

It should by now be clear that free, interacting, and supersymmetric {\it classical} field theories can be manifestly formulated
 independently of choices of local unit systems using Weyl invariance.
This viewpoint clarifies the origins of masses in field theories, particular in curved spaces where it is necessary to survey
all possible couplings to the background geometry and scale (rather than just the compensating mechanism alone) to obtain the theories
we have described here.

The above results are all classical, but in fact the greatest impact of our ideas may be to  quantization.
At the quantum level, scale invariance is anomalous while our classical approach gauges this symmetry and promotes
 curved Riemannian backgrounds $g_{\mu\nu}$ to conformal equivalence classes $[g_{\mu\nu},\sigma]$.
There are strong reasons to believe that this approach can be very fruitful based on ideas coming from the AdS/CFT
correspondence~\cite{Maldacena:1997re,Witten:1998qj,Gubser:1998bc,Aharony:1999ti} where renormalization group flows can be formulated holographically~\cite{de Boer:1999xf} and scale or Weyl anomalies
become geometric~\cite{Henningson:1998gx}. Indeed conformal geometry computations of Poincar\'e metrics~\cite{FG} can be used to obtain 
physical information about these anomalies~\cite{de Boer:1999xf}. At the very least the tractor techniques provide a powerful machinery
for these types of computations, and optimally can provide deep insights into the AdS/CFT correspondence itself.

%Coupling to the scale, we have successfully written spin~1/2 Dirac equation, spin~3/2 Rarita-Schwinger equation, and interacting Wess-Zumino model in a manifestly Weyl Invariant way.  We find special values of the Weyl weights when the scale can be eliminated without sacrificing the Weyl invariance of the underlying theories.  At that special weight, $w=-\frac{d}{2}$, we recover Weyl invariant Dirac operator and the Weyl invariant Rarita-Schwinger equation written by Deser and Nepomechie.  Since our construction unifies the massive and massless theories in a single framework, masses are related to Weyl weights, and Breitenlohner--Freedman bounds  arise naturally as a consequence of the reality of these weights.  

%We also show that the compensating mechanism falls short of mapping the entire landscape of Weyl invariant theories, and to find all possible Weyl invariant theories, we have to naturally rely on tractors.  Our tractor construction naturally produces anti de Sitter theories obtained by log-radial reduction of Minkowski theories in one higher dimension. 

\section{Acknowledgements}
%A.R.G. was supported by Marsden Grant no.\ 06-UOA-029 and a membership
%of the Institute for Advanced Study Princeton.  
%A.W. is indebted to
%the University of Auckland for warm hospitality.
We thank Rod Gover for a partial collaboration in early stages of this work. We also thank Bayram Tekin and Tekin Dereli
for discussions.

\appendix

\section{Doubled Reduction}

\label{Doubled Reduction}

There is a rather explicit relationship between the tractor theories we write down
(when the background is conformally flat) and  log radial reductions from massless theories in $(d + 1)$-dimensional flat Lorentzian spaces to massive ones in $d$-dimensional (anti) de Sitter spaces~\cite{Biswas:2002nk, Hallowell:2005np}. The mathematical underpinning of this relationship is the connection between conformal and projective structures~\cite{BEG}. 
In fact, the independent field content of our tractor models, (which typically inhabit the top and middle slots of tractor fields) 
corresponds precisely to that of these log radial reductions. Therefore, for completeness, in this Appendix we present the
log radial reduction for spinor theories.

For spinor theories there are two possible reduction schemes  depending on how Dirac matrices in adjacent dimensions are handled.
One approach is to write the $(d+1)$-dimensional Dirac matrices as $\Gamma^M = (-i\Gamma^{d+1}\gamma^m,\Gamma^{d+1})$ 
where $\gamma^m$ then obey a $d$-dimensional Clifford algebra. Alternatively, beginning with the $d$-dimensional Dirac matrices
$\gamma^m$, the $(d+1)$-dimensional Dirac matrices are then  built by doubling, namely $\Gamma^M=(\sigma_z\otimes\gamma^m,\sigma_x\otimes \boldsymbol 1)$. Irreducibility of the spinor representations produced in these ways depends on the dimensionality~$d$ and metric signature, but
these details do not concern us here.  Either approach can be related to tractors, but the latter approach (which we adopt here) produces a doubled set of equations 
for which this relationship is simplest--we shall call it a ``doubled reduction''.

We start by writing the flat metric in log radial coordinates 
\bea
ds^{2}_{\rm flat} = dX^M G_{MN} dX^N = e^{2u}(du^2+ ds_{\rm dS} ) = E^A \eta_{AB} E^B \, . \label{logr}
\eea
Here, we consider the case where the underlying manifold is de Sitter for reasons of simplicity (the corresponding AdS computation  is not difficult either).
Notice that the indices $M, N,\ldots$  and $A, B,\ldots$  are not tractor indices but rather $(d + 1)$-dimensional curved and flat ones, respectively,  
while $\mu, \nu,\ldots$  and $m, n \ldots$ are $d$-dimensional. Note that in this background~$\Rho = \frac{d}{2}$. The $(d+1)$-dimensional vielbeine are 
\be
E^5 = e^udu \, , \qquad E^m = e^ue^m \, ,
\ee
where $e^m$ is the $d$-dimensional de Sitter vielbein.
The  de Sitter spin connection $\omega_{mn}$ obeys 
\be
de^m + \omega^{mn} \wedge e_n = 0 \, ,
\ee
in terms of which the $(d + 1)$-dimensional flat spin connection reads 
\be
\Omega^{mn} = \omega^{mn} \,, \qquad \Omega^{5n} = -e^n  \, .
\ee
We use the label $5$ to stand for the reduction direction so that 
$A = (m,5)$ and  $M = (\mu,u)$. The $(d + 1)$-dimensional Dirac matrices and covariant derivative acting on spinors are\bea
\Gamma^A &=&  \Big(  \begin{pmatrix} \gamma^m & 0 \\ 0 & - \gamma^m \end{pmatrix}\, , \,  \begin{pmatrix} 0 & \boldsymbol 1 \\ \boldsymbol 1 &0 \end{pmatrix} \Big) \, ,\nn\\[2mm]
\nabla_M &=& 
%( E^{\mu}{}_m \bar \nabla_{\mu},  E^{u}{}_5\nabla_{u}) = 
\Big( \begin{pmatrix}  \nabla_{\mu} & \frac{1}{2}\gamma_{\mu} \\ -\frac{1}{2}\gamma_{\mu} & \nabla_{\mu} \end{pmatrix} \, , \, \begin{pmatrix} \partial_u&0\\0&\partial_u \end{pmatrix} \Big)\, ,
\eea
where $\nabla_\mu$ on the right hand side of the second line is the de Sitter covariant derivative and $\gamma^m$ are the $d$-dimensional Dirac matrices. For future use we call
\be
\wt \nabla_\mu = \begin{pmatrix}  \nabla_{\mu} & \frac{1}{2}\gamma_{\mu} \\ -\frac{1}{2}\gamma_{\mu} & \nabla_{\mu} \end{pmatrix}\, .
\ee
Acting on spinors $[\wt \nabla_\mu,\wt \nabla_\nu]=0$.
The Dirac conjugate is defined as
$
\overline \Psi = i\Psi^\dagger \Gamma^0$.
We now have enough technology to start the doubled reduction.  As a warm up, we consider a  Dirac spinor
\be
\Psi = \begin{pmatrix}   \psi \\ \chi \end{pmatrix}\, ,
\ee
with action principle 
\be
S_{1/2}= - \int \sqrt{-G}\  d^{d+1}\!X\,  \overline \Psi \Gamma^M \nabla_M \Psi \, .
\ee
Upon making the field redefinition 
\be
\Psi =  e^{-\frac{ud}{2}}  \begin{pmatrix}   \psi \\ \chi \end{pmatrix} \, ,
\ee
all explicit $u$ dependence disappears and the action becomes 
\be
S_{1/2}= - \int du d^d x \sqrt{-g}\,  (\bar \psi, -\bar \chi) \begin{pmatrix} \slashed \nabla & \partial _u \\ \partial_u & - \slashed \nabla    \end{pmatrix}  \begin{pmatrix}   \psi \\ \chi \end{pmatrix} \, .
\ee
Varying this action, defining ({\it a l\'a} Scherk and Schwarz~\cite{Scherk:1979zr}) $$\partial_u = w + \frac{d}{2}\, ,$$ 
and redefining $\chi$ by multiplying it with $\sqrt{2}$,  we recover the pair of equations of motion following from the tractor computation~\eqn{dirac2}
in a canonical choice of scale.
Having warmed up on spin~1/2, our next task is the  spin~3/2 doubled reduction.

\noindent
The $(d+1)$-dimensional  Rarita-Schwinger action is 
\be
S_{3/2}= - \int \sqrt{-G}\,  d^{d+1}\!X \, \overline \Psi_M \Gamma^{MNR} \nabla_N \Psi_R \, .
\ee
Using the  log radial coordinates~\eqn{logr}  and rescaling fields
\be
\Psi_M = e^{-\frac{u(d-2)}{2} } \begin{pmatrix}   \psi_M\\ \chi_M \end{pmatrix} \, ,
\ee
the action becomes devoid of explicit $u$ dependence and implies the doubled set of equations of motion
%\be
%S_{3/2}= - \int \sqrt{-g}\,du d^d x \, [\overline \Psi_u   \Gamma^{5\nu\rho} \bar \nabla_{\nu} \Psi_{\rho} +  \overline \Psi_{\mu}  (  \Gamma^{\mu\nu\rho} \bar \nabla_{\nu} \Psi_{\rho}  +  \Gamma^{\mu 5\rho}  (\partial_u-\frac{d}{2}+1) \Psi_{\rho}  +   \Gamma^{\mu\nu5} \bar \nabla_{\nu} \Psi_{u}) ]\, .
%\ee
\be
\Big[
\begin{pmatrix} \gamma^{\mu\nu\rho}&0\\0&-\gamma^{\mu\nu\rho}\end{pmatrix}
 \wt \nabla_{\nu} 
 -  \begin{pmatrix}0& \gamma^{\mu\rho}\\\gamma^{\mu\rho}&0\end{pmatrix} (\partial_u-\frac{d}{2}+1)
  \Big]\begin{pmatrix} \psi_\rho\\ \chi_\rho\end{pmatrix} = 0\, .
\ee
(Note that the $\Psi_u$ variation simply implies the constraint that is a consequence of the above equation.)
Redefining the $\chi$ field as before, and equating $\partial_{\mu} =w + \frac{d}{2}$, the above equation agrees with the tractor Rarita-Schwinger~\eqn{rarita} explicated in the canonical choice of scale.

\end{document}